\documentclass[10pt]{blois07}
\usepackage{graphicx}
\usepackage{cite,./mcite}

\begin{document}
\title{Blois07/EDS07 Proceedings \\ RHIC physics:  short overview}
\author{Anna Sta\'sto}
\institute{Penn State University, 104 Davey Lab., University Park, 16802 PA, USA}
\maketitle

\begin{abstract}
This talk gives a very short overview of some of  the important  physical phenomena observed at the Relativistic Heavy Ion Collider (RHIC). The emphasis is put on the multiplicities, hard probes and the properties of the initial state.
\end{abstract}

%%%%%%%%%%%%%%%%%%%%%%%%%%%%%%%%%
\section{Introduction}
\label{sec:intro}

\begin{figure}[t]
\centerline{\includegraphics[width=7.5cm]{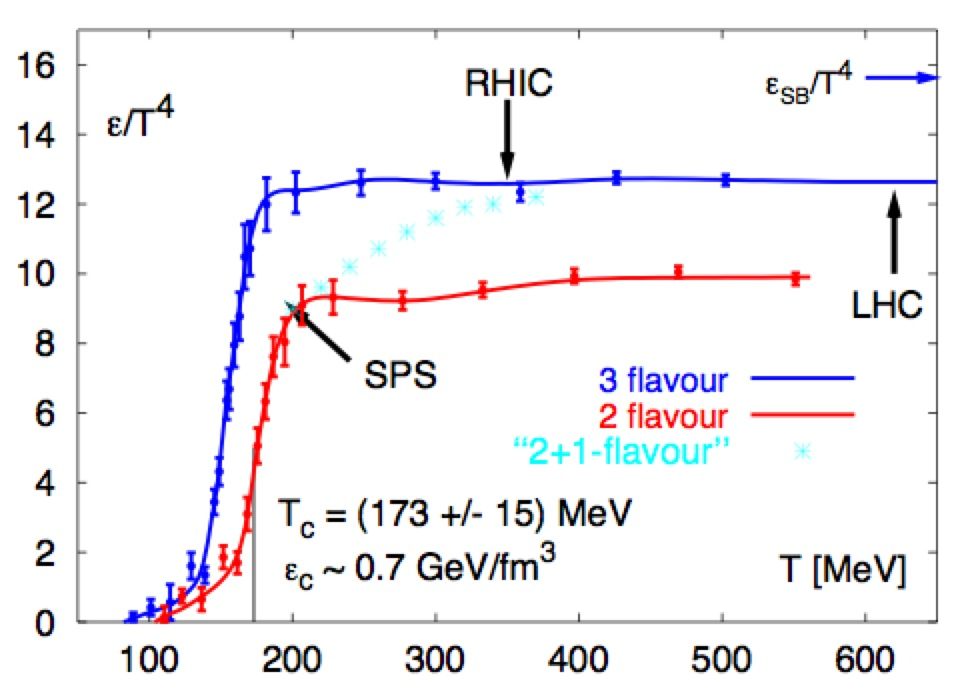}}
\caption{Energy density $/T^4$ as a function of the temperature. The blue arrow with $\epsilon/T^4$ is the Stefan-Boltzman limit. Figure by F.~Karsch \protect\cite{Karsch}.}
\label{karsch}
\end{figure}

\begin{figure}[h]
\centerline{\includegraphics[width=7.5cm]{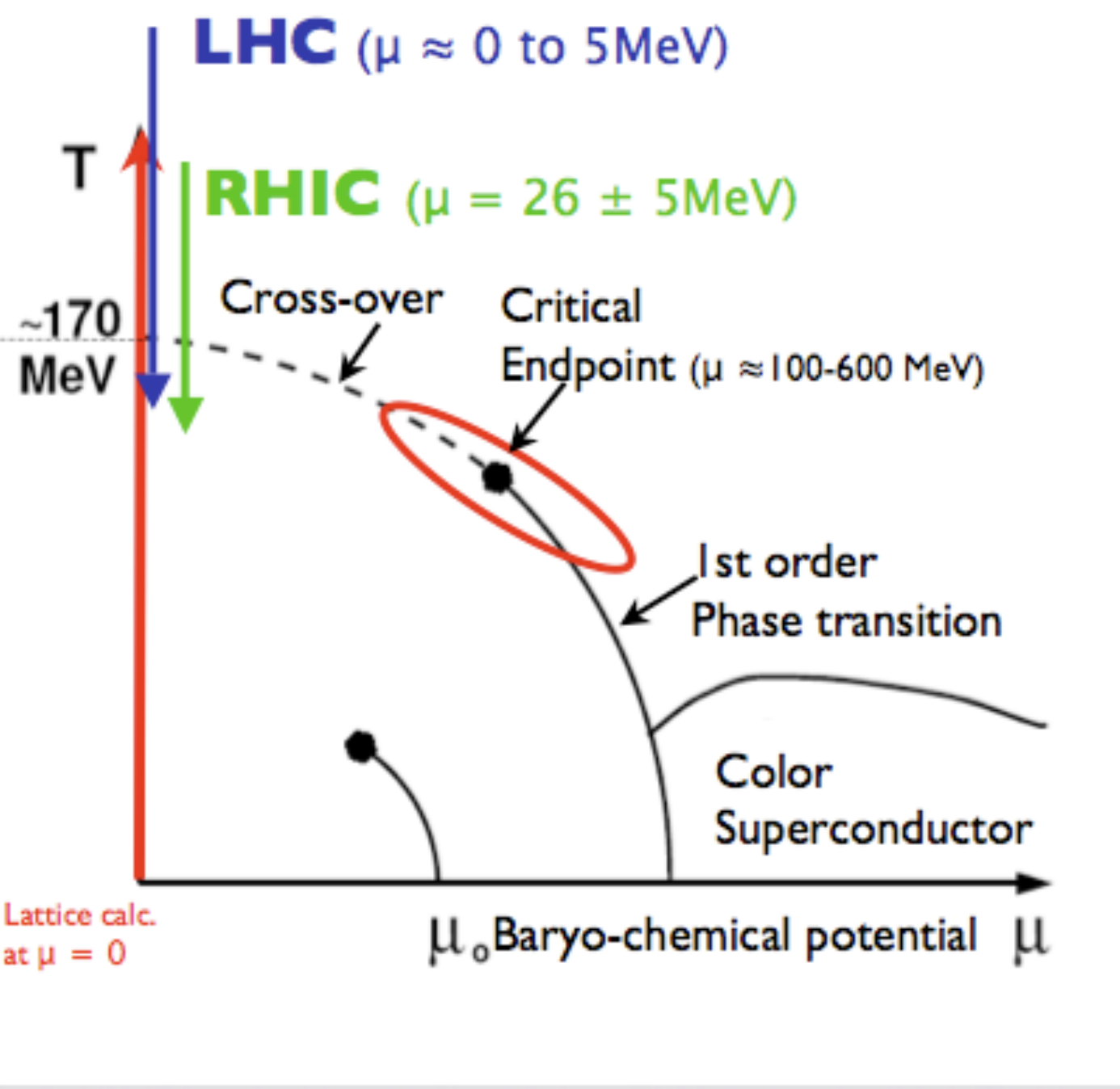}}
\caption{Phase diagram of QCD.}
\label{phase_diagram}
\end{figure}

The Relativistic Heavy Ion Collider (RHIC) is a unique machine designed to create a very high energy density over an extended region as a result of nuclei collisions. This process  enables to  investigate  the  collective phenomena in Quantum Chromodynamics.  In particular, one hopes to create  the quark-gluon plasma which is the state of deconfined quarks and gluons. 
According to cosmology such  state  existed at the very early Universe.  RHIC machine is capable of accelerating and colliding different hadronic systems: proton-proton, nucleus-nucleus, deutron-nucleus at a wide range of centre-of-mass energies, from $19.6$ to $200 \; {\rm GeV}$ per nucleon pair.   Just after the collision, a very  high quantity  of quarks and gluons is created. It is estimated that, the formation time for the  initial density is about $0.35\;  {\rm fm/c}$.  After that, the system  thermalizes very quickly and reaches thermodynamic equillibrium.  Estimates suggest that this happens rather quickly,  with very short thermalization times of the order of $1 \; {\rm fm/c}$.  The assumption of thermalization  is vital for the application of the hydrodynamics \cite{Heinz:2004ar} which is used to describe the expansion of the system. 
When the system expands and cools down, the quarks and gluons form  hadrons which finally reach the detectors.  The phase transition from the quark-gluon plasma to hadrons occurs at rather large value of  strong  coupling which is beyond the applicability of the perturbative methods.  Nevertheless, one can explore thermodynamic properties of QCD using lattice methods, see for example \cite{Karsch:2006sf}. In Fig.~\ref{karsch} we show the result of the lattice calculations \cite{Karsch:2006sf} for the energy density divided by $T^4$ as a function of temperature. The results clearly show the phase transition at the critical temperature  $T_c$ of about $173\; {\rm MeV
}$. The critical energy density corresponding to this temperature is about $\epsilon_c \simeq 0.7 \;  {\rm GeV/fm^3}$. The energy density at RHIC \cite{BNL:whitepaper} can be calculated from the transverse energy density at midrapidity via Bjorken formula: $\langle \epsilon \rangle \simeq \frac{1}{\tau A} \frac{dE_T}{dy}$, where
$A$ is the overlap area for the colliding nuclei. The average energy density depends crucially on the estimates of the  time $\tau$ at which it is evaluated. For thermalization times in the range $0.6-1.0 \; {\rm fm/c}$ the  average energy density is about $9.0-5.4 \; {\rm GeV/fm^3}$. This is well above the critical density obtained from lattice calculations, compare Fig.~\ref{karsch}. It is interesting to note that, the energy density of the cold nuclear matter is  about $0.15 \; {\rm GeV/fm^3}$. In Fig.~\ref{karsch}  the energy densities probed by RHIC and LHC are also indicated. Notably, the calculations signal that  already at SPS energies, the transition from hadron phase to quark-gluon plasma occured. Lattice calculations enable to probe  the phase diagram of QCD, shown schematically in Fig.~\ref{phase_diagram}.
The vertical axis is the temperature  $T$ and the horizontal one is the baryo-chemical potential.

%%%%%%%%%%%%%%%%%%%%%%%%%%%%%%%%%
%%%%%%%%%%%%%%%%%%%%%%%%%%%%%%%%%

The region above the dashed-solid lines is the quark-gluon plasma phase, whereas the region below,
at small temperatures and baryo-chemical potential, is the hadron phase. The transition between the two is a smooth crossover. Both RHIC and LHC probe the quark-gluon phase and the transition region
at small values of the baryon chemical potential. Further to the right, for higher values of the baryo-chemical potential the critical endpoint is expected and the transition becomes of the 1st order. 
At high values of the baryo-chemical potential and smaller temperatures, new phase appears, the color superconductor. This phase is not accessible at the high energy colliders.

%%%%%%%%%%%%%%%%%%%%%%%%%%%%%%%%%
%%%%%%%%%%%%%%%%%%%%%%%%%%%%%%%%%

The fact that  RHIC  reached quark-gluon plasma domain in the nucleus-nucleus collisions suggest that, at least in some of the measured observables, one can expect to   see dramatic effects as compared with proton-proton collisions. In reality the situation is quite complicated. Indeed, large differences are seen, for example  the suppression of  $p_T$ distributions at large values of the transverse momenta with respect to the (scaled) proton-proton collisions.
On the other hand, bulk properties, like total multiplicities and  rapidity distributions,  have quite similar shapes and energy dependences
as compared with the scaled proton-proton measurements. We are going to review some of the observations
performed at RHIC and discuss the phenomenological descriptions.

%%%%%%%%%%%%%%%%%%%%%%%%%%%%%%%%%
\section{Multiplicities}
\label{sec:mult}

\begin{figure}[h]
\centerline{\includegraphics[width=8.5cm]{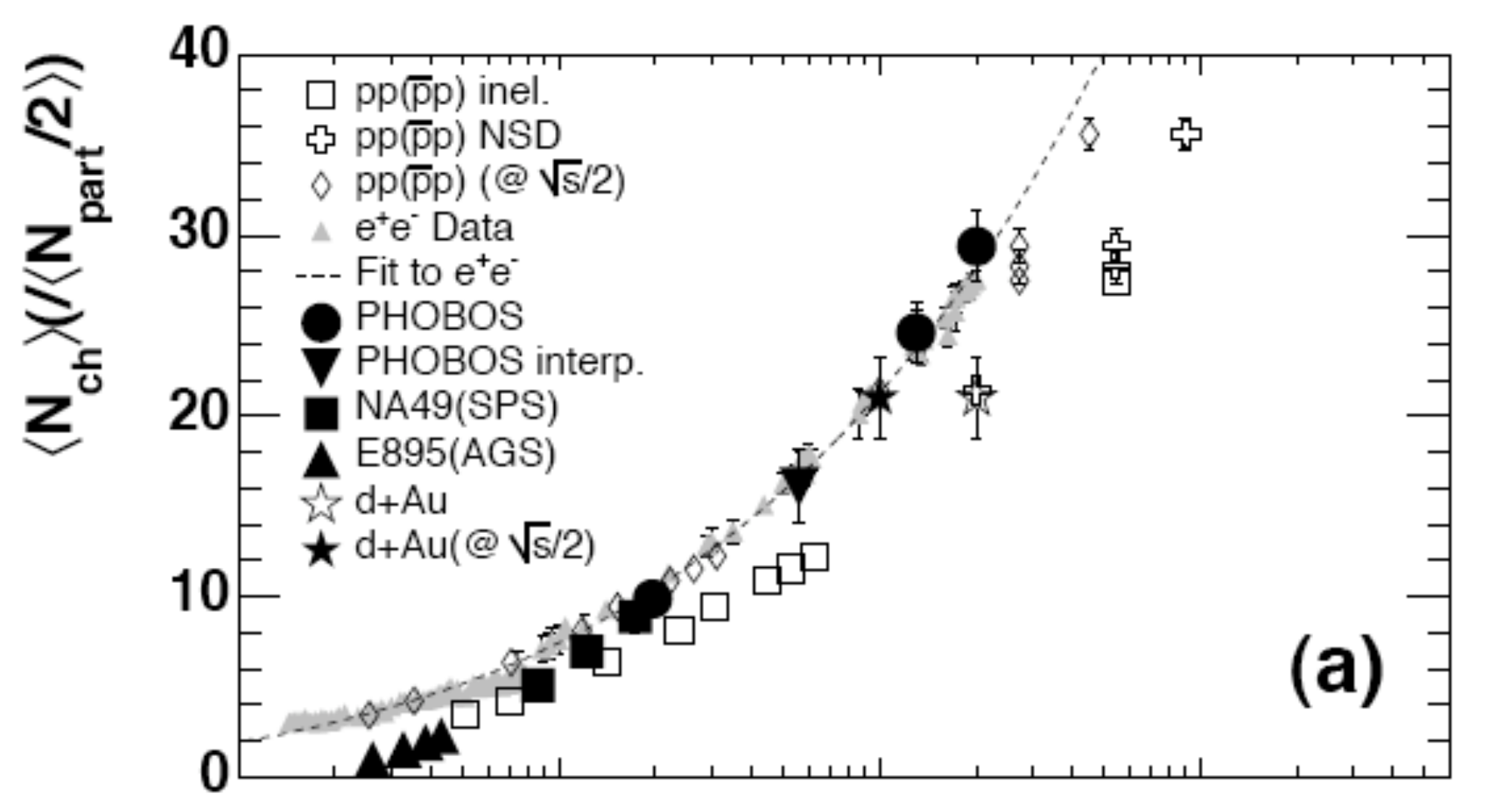}}
\caption{Total multiplicity scaled by the number of participating pairs as a function of the energy.
Compilation of data from $e^+e^-$, $pp(p\bar{p})$, $AA$ and $dA$ collisions. Figure by PHOBOS collaboration from \protect\cite{BNL:whitepaper}.}
\label{tot_mult}
\end{figure}

In Fig.~\ref{tot_mult} from \cite{BNL:whitepaper} the measurements of the  total multiplicity in $AA$, $pp(p\bar{p})$, $e^+e^-$ collisions, scaled by the number of participating nucleon pairs $N_{\rm part}/2$ are shown as a function of the increasing energy. The smooth rise with energy  is well described by the $\ln^2 s$ behavior over wide range of the energies. What is striking, is the fact that the data for scaled multiplicity for nucleus-nucleus collisions  lie on top
of the data points for $e^+e^-$ collisions. Proton-proton data lie lower, most probably due to the leading particle effect.  In the $pp$($p\bar{p}$) collisions, lots of the energy is taken by the quark spectators into the forward region, and only a fraction of the energy is used for the production of the secondary particles. The proton-proton data can be superimposed onto the nucleus-nucleus and $e^+e^-$ data when the energy is rescaled
by a factor of $1/2$. This universality of the multiplicites indicates that the bulk of the produced particles  depends only on the total energy (and $N_{\rm part}$) and not the species of the colliding particles.

%%%%%%%%%%%%%%%%%%%%%%%%%%%%%%%%%
%%%%%%%%%%%%%%%%%%%%%%%%%%%%%%%%%

The measured  multiplicity at mid-(pseudo)rapidity in nucleus-nucleus collisions increases slower with energy , as $\ln s$. This simple linear in $\ln s$ extrapolation to the LHC energies predicts that $\frac{2}{N_{\rm part}}\frac{dN}{d\eta}|_{\eta=0} \simeq 6-7$. Various theoretical predictions for LHC energy are larger than this simple estimate and   span a wide range, up to nearly $40$ for the value at midrapidity.  This is probably connected to the fact that in most of these calculations some semi-perturbative component is present which results in a power-like increase of the multiplicity rather than the logarithm. This large uncertainty of the phenomenological extrapolations reflects our rather limited theoretical understanding of the energy dependence of multiplicities.

%%%%%%%%%%%%%%%%%%%%%%%%%%%%%%%%%
\subsection{Extended longitudinal scaling}
\label{sec:extlongscal}

PHOBOS collaboration performed measurements of the rapidity distributions for various energies (and systems) and found that the distributions exhibit limiting fragmentation property which is also called extended longitudinal scaling \cite{Back:2002wb}. This means that when viewed in the rest frame of one of the projectiles the (pseudo)rapidity distribution becomes a function of only $\eta'=\eta-Y_{\rm beam}$  where $Y_{\rm beam}$ is the rapidity of the beam.  Therefore  the rapidity distribution
 in the regime around $\eta' \sim 0$   is dominated by the fragments of the broken target whereas the fragments of the projectile move with increasing velocity as the energy is further increased (to study these fragments one has to go to the rest frame of the projectile $\eta+Y_{\rm beam}$). The limiting fragmentation also requires that the interaction between the target and the projectile does not depend appreciably on the energy. This scaling means that the rapidity distributions must be determined very early in the collision,
most probably  by the initial state.  

%%%%%%%%%%%%%%%%%%%%%%%%%%%%%%%%%
\begin{figure}[h]
\centerline{\includegraphics[width=7.5cm]{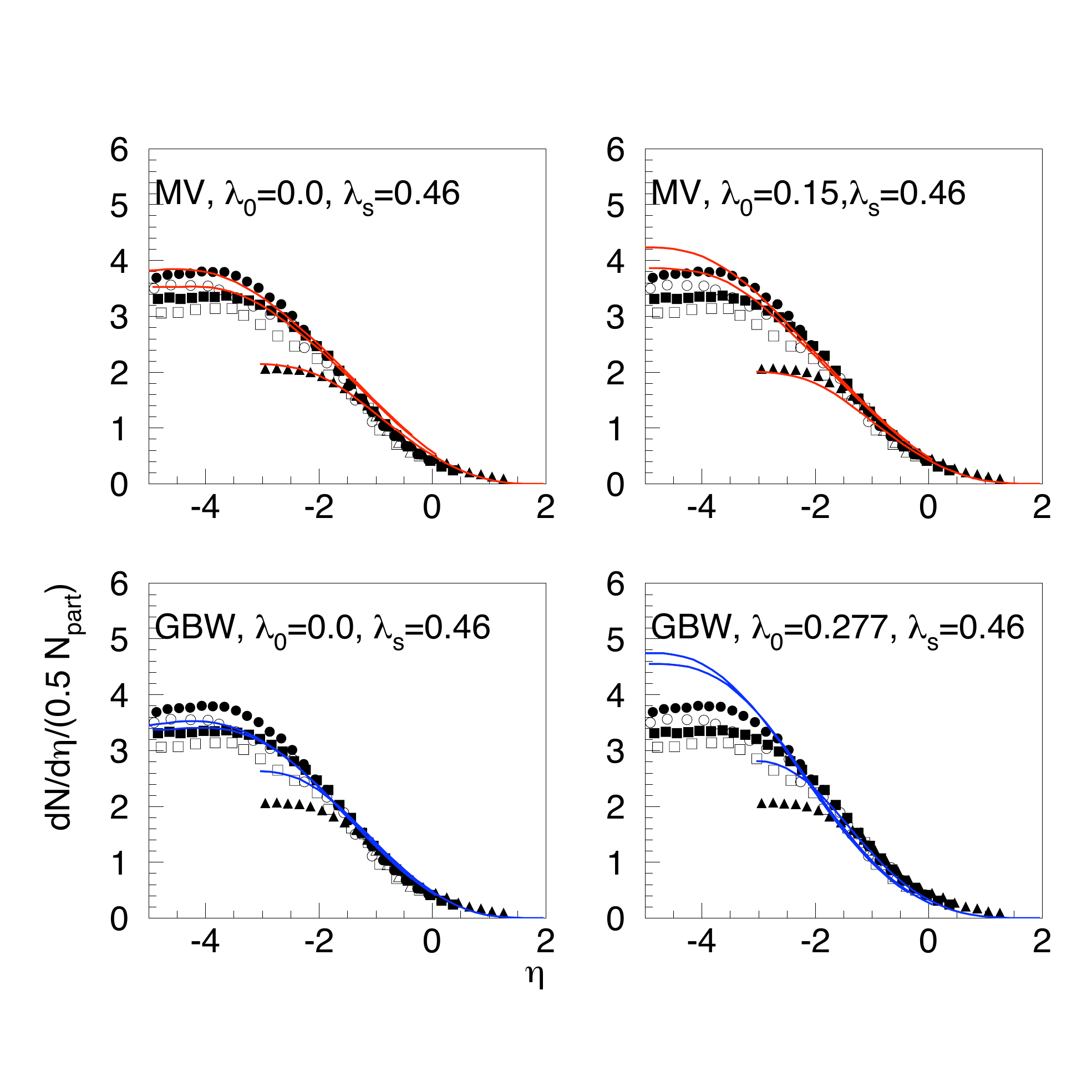}}
\caption{Rapidity distributions for the nucleus-nucleus collisions as a function of the shifted variable $\eta'=\eta-Y_{\rm beam}$ for various center-of-mass energies. Curves are from calculation using the CGC nonlinear equations \protect\cite{Gelis:2006tb}.}
\label{rap_distr}
\end{figure}
%%%%%%%%%%%%%%%%%%%%%%%%%%%%%%%%%

Fig.~\ref{rap_distr} illustrates this phenomenon. The curves shown in figure are obtained from the calculation \cite{Gelis:2006tb} based on the $k_T$-factorization approach together with the evolution via nonlinear equation for the gluon density. In this approach the rapidity distribution is evaluated as the convolution of two parton densities corresponding to the target and the projectile respectively. In the fragmentation region, the target parton density is evaluated at rather large values of Bjorken $x$ whereas the projectile density at rather small values of $x$. The essential part of this calculation, which enables to reproduce the observed scaling, is the fact that the rapidity distribution in the target fragmentation region is dominated by the initial state parton density of the target, probed at large values of the Bjorken variable $x$. At these values the parton density possesses Bjorken scaling,i.e. does not depend on the scale at which it is evaluated. The projectile density, which depends on rather small values of  $x$ and small scales, is saturated since it is evaluated from the evolution equation which takes into account nonlinearities important at high parton  densities. Therefore  this density does not depend much on the $x$ values and hence the center of mass energy. We see that even though the approach is semi-perturbative it does capture the  essential physics features necessary to reproduce the limiting fragmentation phenomenon.

%%%%%%%%%%%%%%%%%%%%%%%%%%%%
\section{Hard probes}
\label{sec:hard_probes}

\begin{figure}[h]
\centerline{\includegraphics[width=7.5cm]{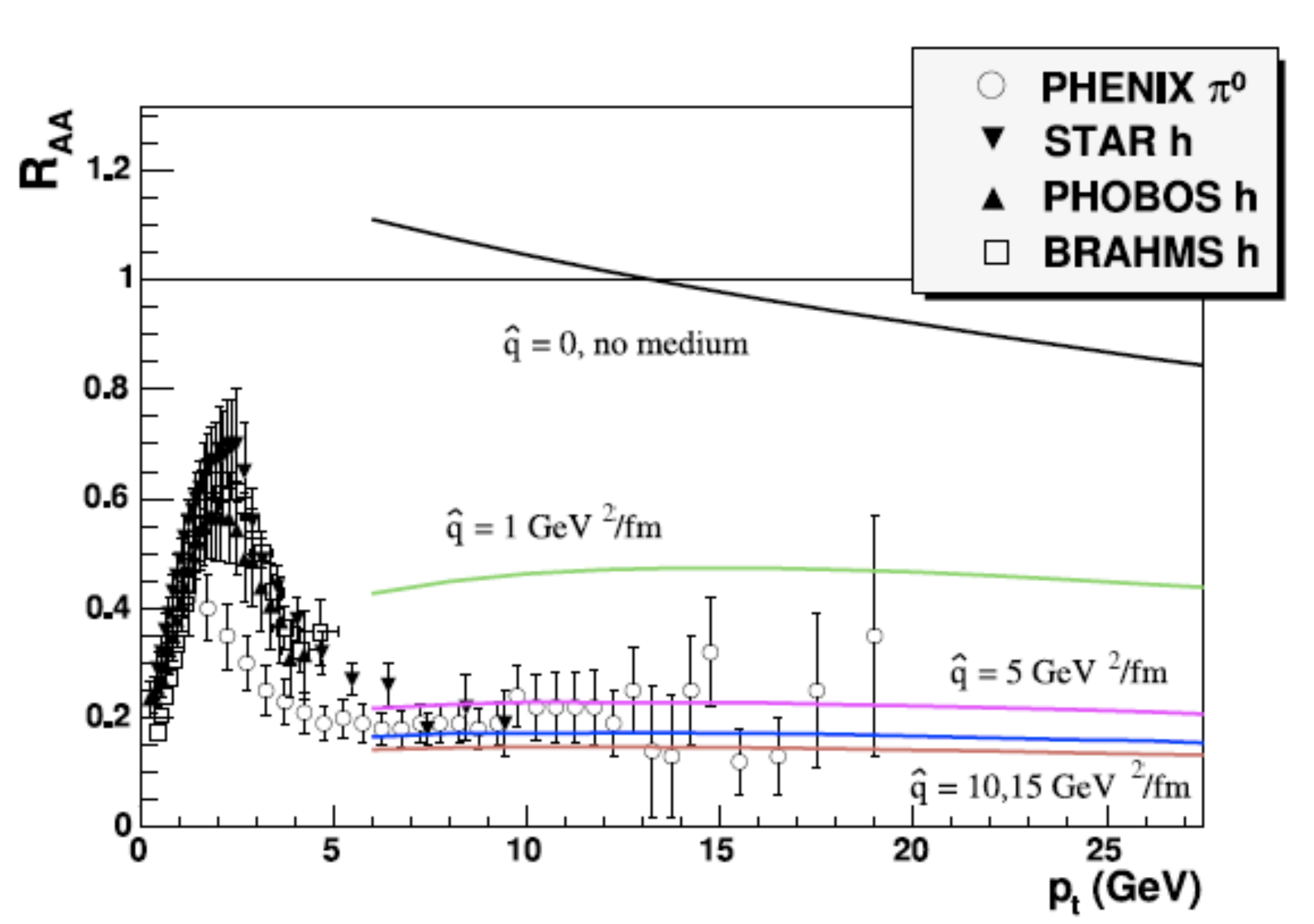}}
\caption{$R_{AA}$ for the produced hadrons as a function of their transverse momentum. Figure from \protect\cite{Salgado:2007rs}.}
\label{pt_had_raa}
\end{figure}

To explore the properties of the quark-gluon plasma created in the high energy nucleus-nucleus collisions 
 experiments at RHIC  measured the suppression of the production of the  high $p_T$ particles
and heavy quarks. These are excellent probes of the produced medium, and due to the large
difference in scales (high $p_T$ as compared to the bulk of low $p_T$ particles ), it should
be in principle possible to employ the perturbative methods.  One usually quantifies the effect of the medium by evaluating the ratio $R_{AA}=\frac{\sigma_{pp}^{inel}}{\langle N_{coll}\rangle}\frac{d^2 N_{AA}/d\eta dp_t}{d^2 \sigma_{pp}/d\eta dp_t}$. 
As seen from Fig.~\ref{pt_had_raa} from \cite{Salgado:2007rs} the $R_{AA}$ ratio  is significantly below $1$ even at the very high values of $p_T$.
This phenomenon is called the jet quenching  which is interpreted as the interaction   of the produced jet (or rather leading
high $p_T$ particle) with the produced medium. The interaction results in the significant energy loss of the leading high $p_T$ particle.  In fact the effect is so large that it indicates that the high $p_T$ particles which reach the detectors are emitted from a relatively thin outer shell of the high density region.
%%%%%%%%%%%%%%%%%%%%%%%%%%%%%%%%%
%%%%%%%%%%%%%%%%%%%%%%%%%%%%%%%%%

 To describe the jet quenching one usually starts from the standard collinear factorization formula for the exclusive  hadron production in the vacuum
\begin{equation}
\sigma_{\rm vac}^{AB\rightarrow h}= f_A(x_1,Q^2)\otimes f_B(x_2,Q^2) \otimes \hat{\sigma}(x_1,x_2,Q^2) \otimes D^{f\rightarrow h} +{\cal O}(\Lambda/Q) \; ,
\label{eq:coll_vac}
\end{equation}
where $f_A,f_B$ are parton distribution functions evaluated at scale $Q^2$ , $\hat{\sigma}$ is the partonic cross section and 
$D^{f\rightarrow h}$ is the  fragmentation function from parton $f$ to hadron $h$.
To evaluate the production in the medium one convolutes the above expression with the so called quenching weight \cite{Salgado:2003gb}
\begin{equation}
\sigma_{\rm med}^{AB\rightarrow h}=\sigma_{\rm vac}^{AB\rightarrow h} \otimes P(\Delta E,L,\hat{q}) \; ,
\label{eq:coll_med}
\end{equation}
which depends on the length of the  medium and the parameter $\hat{q}$ which acts as a transport coefficient for the medium.
One has to emphasize that although in the proton-proton collisions the factorization formula (\ref{eq:coll_vac}) is well established, (up to higher twist corrections)
it is not proven to hold in form (\ref{eq:coll_med}) for  collisions which involve large nuclei. Final state interactions could in principle affect the partonic cross section $\hat{\sigma}$ in  a non-factorizable way. It is therefore an assumption, that after the partons have been produced in the hard scattering process, one can further factorize their interactions into one universal function $P$.
The curves in Fig.~\ref{pt_had_raa} are from  \cite{} which is calculation based on the theoretical framework described above. The quenching parameter $\hat{q}$ was fitted to get the best description of the data.

The production of the heavy quarks is another excellent process. In principle the calculation  framework is the same as above.
One does expect  some differences due to the fact that in the vacuum the radiation patter for heavy quarks is different from the light quarks.
Due to the their mass heavy quarks should radiate less than the light quarks at angles smaller than $\theta_0=m/E$ where $m$ is the mass of the heavy quark and $E$ is its energy. This is called the dead cone effect \cite{Dokshitzer:2001zm}.  Therefore the net effect would be less supression 
(larger $R_{AA}$) than for the light quarks. 
%The measurements together with the theoretical calculations are shown in Fig.~\ref{pt_hq_raa} from \cite{Armesto:2005mz}.
The measured suppression for the heavy quarks  is of the same order  as for the light quarks, and the theoretical predictions (for example \cite{Armesto:2005mz}) do not quite predict such a large value.  One has to  emphasize that  the heavy quark production in $pp$ (from  electrons emerging from semi-leptonic decays) is underestimated by the NLO perturbative QCD calculation  by a factor of about $2$ for PHENIX data  and by about factor $5$ for the STAR data.
Therefore the process with the heavy quarks  calls for  better  understanding, possible taking into account  various  effects : collisional energy loss, bottom/charm ratio or factorization breaking.

We conclude our discussion of the hard probes with the description of the new calculational methods
which employ the AdS/CFT correspondence to evaluate the jet quenching parameter.
In this approach \cite{Liu:2006he}, the expectation value of the  Wilson loop,  as an average over the medium is evaluated using the string/gauge duality in the limit of the strong coupling constant. This expectation value is 
 obtained  as an exponential of the string action  evaluated at the minimum
$$
\langle W^F(C)\rangle = \exp(i S(C)-i S_0) \; ,
$$
where $S$ is the Nambu-Goto action with the metric on the $4+1$ dimensional AdS space.
In the case of the finite temperature, the corresponding metric is that of the AdS Schwarzschild black hole. In the strong coupling and the multicolor limit the problem becomes classical, i.e. reduces to  finding the extremum of the action. It is possible, using this method, to evaluate  the value of the jet quenching parameter which in this case is
$\hat{q}_{SYM}=5 \; {\rm GeV^2/fm}$.  We note however, that all these results can be derived only
for the case of the N=4 SYM theory.
%%%%%%%%%%%%%%%%%%%%%%%%%%%%%%%%%
\section{Initial state}
\label{sec:initial_state}

The processes described above clearly indicate the presence of the dense medium in the final state. 
The natural question arises whether one can also observe the effects coming from the initial state, namely the wave function of the colliding nuclei. BRAHMS collaboration performed a measurement of the high $p_T$ suppression as a function of rapidity for $dA$ collisions \cite{Arsene:2004ux}.
Whereas at midrapidity no suppression is observed for this process, the $R_{CP}$ clearly shows a decreasing trend when moving into forward rapidity. This phenomenon was quite successfully described by the models which involve saturation effects in the gluon density
or in general by the Color Glass Condensate model \cite{Iancu:2003xm}. In  CGC the basic object is the wave function of the nucleus. Obviously the complete knowledge of such wave function requires methods which go beyond that known in preturbation theory. One can nevertheless
calculate the variation of this wave function with energy. This is governed by the renormalization group equation which can be derived
from the Feynman graphs in the leading logarithmic approximation. At very small values of Bjorken $x$ one expects the fast growth
of the gluon density within the nucleus. CGC model together with the renormalization group equations predicts that this growth should be tamed whenever the $x$ becomes sufficiently small. The transition between the fast growth and the regime where the recombination
effects for the gluons become important is governed by the saturation scale $Q_s$ which is a function of the Bjorken $x$.
Thus the saturation scale provides with a dynamical cutoff at low values of $x$ and at low scales.
It is the prediction of the CGC model that the $R_{CP}$ ratio should decrease at  forward rapidities \protect\cite{Kharzeev:2004yx}. CGC model has been successfully
used to describe various observables in heavy ion collisions: multiplicites, rapidity distributions (mentioned already in the previous section) \cite{Kharzeev:2001gp}
and also $R_{CP}$ ratio. One has to emphasize though that there are several questions concerning strict applicability of this approach to the  RHIC data. The values of $x$ are not very small for the RHIC kinematics, the formalism correctly incorporates only gluons
and it has been so far only used at leading order whereas higher orders are known to be very large. Nevertheless,
the CGC approach, mostly due to its interesting properties in the infrared regime, remains  a very attractive approach  both theoretically and phenomenologically  and its predictions should be further confronted with the experimental data.

%%%%%%%%%%%%%%%%%%%%%%%%%%%%%%%%%
\section*{Conclusions}
We have discussed a selection of the phenomena measured at the Relativistic Heavy Ion collider. Clearly, due to our space and time
limitations the list presented here is by no means exhaustive. The bulk properties as shown in the measurements of the multiplicities
are  very similar to that measured in  the simpler systems: proton-proton or even in $e^+e^-$ collisions. The extended longitudinal scaling
of rapidity distributions indicate the importance of the initial state. Hard probes in form of the high $p_T$ particles or heavy quarks
signify the presence of the strongly interacting medium. This is further corroborated by the observation of the strong elliptic flow
in peripheral collisions.  The theoretical descriptions based on hydrodynamics  have been quite successful in describing the hadron spectra
and the anisotropy. Also calculations which  employ the perturbative methods supplemented by the rescattering or recombination(saturation) effects
are able to describe the bulk of the data at large values of the transverse momenta. Nevertheless, despite these incontrovertible successes in phenomenology,  the RHIC data still constitute a significant challenge  
for a theory and call for  a more complete and  coherent  description within QCD.

\section*{Acknowledgments}
I would like to thank the organizers of EDS2007 for the kind invitation to this  very interesting workshop.
This research was supported by the 
U.S. D.O.E. under grant number DE-FG02-90ER-40577 and by the Polish Committee for Scientific Research under  grant No.\ KBN 1 P03B 028 28.

%------------------------------------------------------------------------------
%       Bibliography
%------------------------------------------------------------------------------
\begin{footnotesize}
\bibliographystyle{blois07} 
{\raggedright
\bibliography{blois07}
}
\end{footnotesize}
\end{document}